\documentclass[%
 reprint,
 amsmath,amssymb
 aps,
]{revtex4-1}
\usepackage{graphicx}
\usepackage{dcolumn}
\usepackage{bm}
\usepackage{color}
\usepackage{wasysym}
\usepackage{epstopdf}
\usepackage{verbatim}
\usepackage{natbib}
\usepackage{epsfig}
\usepackage{subfigure}

\begin{document}

\preprint{APS/123-QED}

\title{Dark-state sideband cooling in an atomic ensemble}

\author{Chang Huang}
\author{Shijie Chai}
\author{Shau-Yu Lan}
 \email{sylan@ntu.edu.sg}
 \affiliation{Division of Physics and Applied Physics, School of Physical and Mathematical Sciences, Nanyang Technological University, Singapore 637371, Singapore}




\date{\today}

\begin{abstract}
We utilize the dark-state in a $\Lambda$-type three-level system to cool an ensemble of $^{85}$Rb atoms in an optical lattice [Morigi \emph{et al}., Phys. Rev. Lett. \textbf{85}, 4458 (2000)]. The common suppression of the carrier transition of atoms with different vibrational frequencies allows them to reach a sub-recoil temperature of 100 nK after being released from the optical lattice. A nearly zero vibrational quantum number is determined from the time-of-flight measurements and adiabatic expansion process. The features of sideband cooling are examined in various parameter spaces. Our results show that dark-state sideband cooling is a simple and compelling method for preparing a large ensemble of atoms into their vibrational ground state of a harmonic potential and can be generalized to different species of atoms and molecules for studying ultra-cold physics that demands recoil temperature and below.

\end{abstract}

\pacs{Valid PACS appear here}
\maketitle

Cooling of atoms is indispensable in many quantum science experiments \cite{Met}. To achieve an extremely low temperature of atoms, various stages of cooling mechanisms have to be employed. Doppler and sub-Doppler cooling through lasers are conventionally used to remove the kinetic energy of atoms from room temperature down to the few hundreds or tens of micro-kelvin range. To reach tens of nano-kelvin or an even lower temperature, evaporative cooling \cite{And} and delta-kick cooling \cite{Kov} are the workhorses whose cooling efficiencies, however, are largely affected by the initial temperature. To bridge the gap, free-space Raman cooling \cite{Kase}, Raman sideband cooling \cite{Per,Ham}, or degenerate Raman sideband cooling \cite{Vul,Ker} can be deployed to prepare atoms in the few hundreds of nano-kelvin range. Nevertheless, these methods require tuning the laser frequencies or external magnetic fields to match the vibrational levels to minimize the heating process. For a sizable atomic ensemble in a harmonic potential with a distribution of vibrational frequencies over different lattice sites, simultaneously attaining a high cooling efficiency and short cooling time is challenging.

Alternatively, dark-state sideband cooling offers a wider cooling bandwidth by tailoring the cooling laser absorption spectrum through quantum interference. The dark-state has been utilized in cooling atoms to sub-recoil temperatures in terms of velocity selective coherent population trapping \cite{Asp}. However, it is restricted to atoms that have $F=1$ and $F'=1$ energy levels. Combined with polarization gradient cooling (PGC) to form gray molasses cooling, the dark-state has been used to cool the temperature of atoms down to the few micro-kelvin range. For trapped atoms, the dark-state has been applied in cooling a chain of trapped ions with different vibrational frequencies to their vibrational ground state within 1 ms \cite{Mor,Roo,Mor2,Lin,Lec,Sch,Jor,Qia,Fen}. For an ensemble of neutral atoms in an optical lattice \cite{Kam,Hal,Edg}, cooling to the ground state as well as sub-recoil temperature using the dark-state has not been demonstrated. Here, we realize dark-state cooling of an atomic ensemble in an optical lattice to the vibrational ground state. With the aid of adiabatic cooling, we have observed a sub-recoil temperature of atoms at 100 nK, or 0.27$T_{\textrm{r}}$, after releasing them from the optical lattice, where $T_{\textrm{r}}$ is the recoil temperature.

\begin{figure}[h]
\subfigure{\label{fig:1a}\includegraphics[scale=0.085]{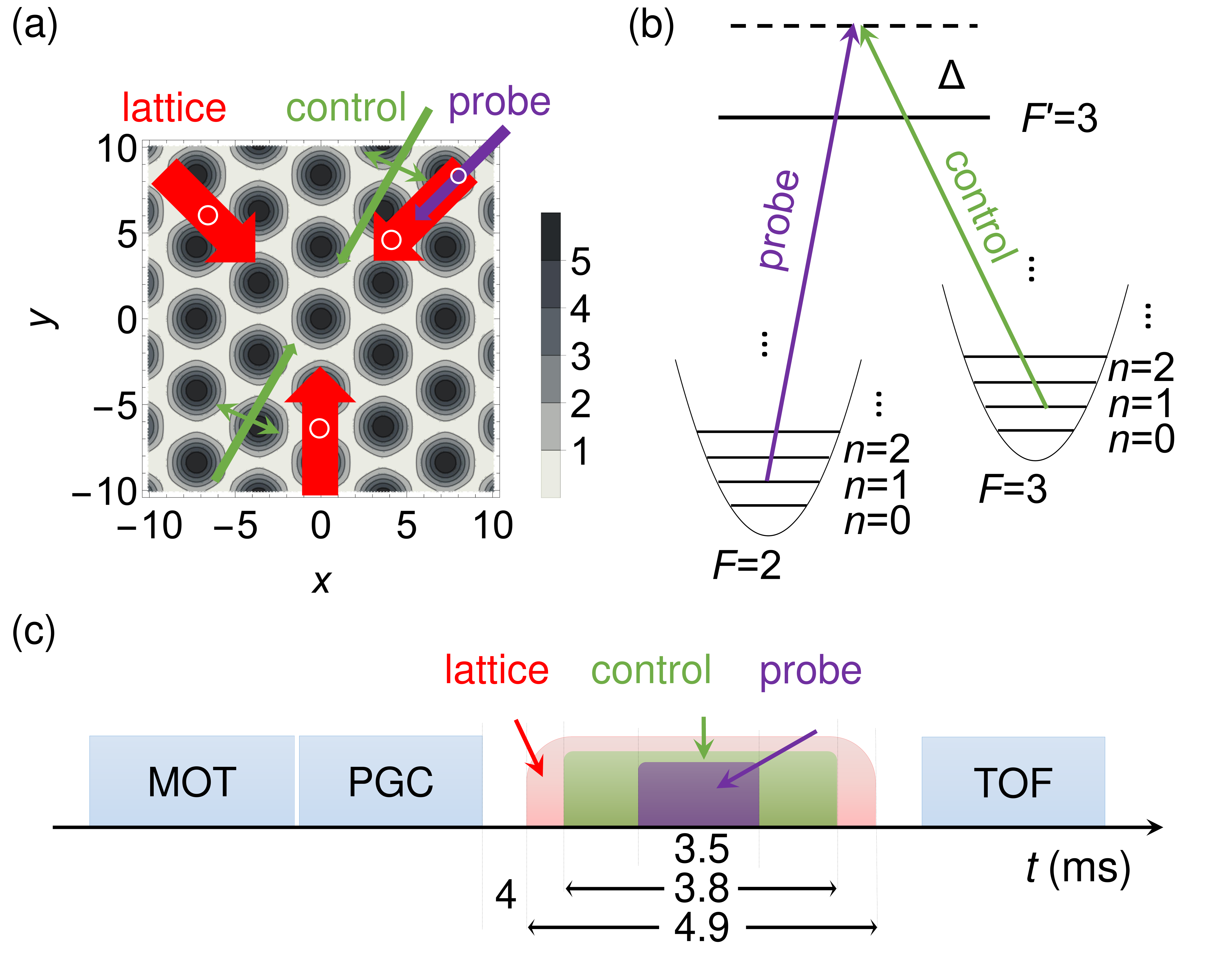}}
\caption{Experimental details. (a) Configuration of the optical lattice and cooling beams. All the laser beams are aligned in the horizonal plane. The white circles indicate the polarization of the laser beams perpendicular to the $x$-$y$ plane. The double arrows represent the polarization of the laser beams lying on the $x$-$y$ plane. The lattice potential is in units of $u$. The $x$ and $y$ coordinates are in units of $k^{-1}$. (b) Relevant energy levels. The control beam modifies the absorption spectrum of the probe beam such that the cooling transitions ($n$ to $n'< n$) dominate the heating transitions ($n$ to $n'\geq n$) when the frequencies of the two beams are adjusted to the two-photon resonance condition.  (c) Timing sequence for dark-state sideband cooling (not to scale).}
\label{fig:1}
\end{figure}

The conventional sideband cooling depends upon the competition between the absorption of the cooling beam from the vibrational level $n$ to $n'<n$ (the cooling transitions or red sidebands) and $n$ to $n'\geq n$ (blue sidebands and carrier transitions), where $n$ is the vibrational quantum number. This is usually done by tuning the cooling beam frequency resonant on the $n$ to $n-1$ transitions in the Lamb-Dicke regime. As a result, for one set of laser parameters, only atoms in the site with one specific trapping frequency can be cooled efficiently. In addition, there are also non-negligible $n$ to $n'=n$ carrier transitions, which limits the cooling duration and efficiency.

In dark-state cooling, two cooling beams, which we denote as the control and probe beams here, couple two narrow linewidth ground states to an excited state separately. The control beam tailors the absorption spectrum of the probe beam such that the $n$ to $n'\geq n$ transitions are greatly suppressed, while the transitions from $n$ to $n'<n$ remain allowed. In the dressed-state picture, the three-level system and two cooling lasers form the condition of coherent population trapping (CPT)\cite{Lou}. The eigenstates of CPT consist of one dark-state and two bright-states. The dark-state corresponds to the carrier transitions, and the energy shift of one of the bright states from the dark-state can be adjusted to match the red sideband transitions which are shifted from the carrier transitions by \cite{Mor2}

\begin{equation}
\delta=\frac{1}{2}(\sqrt{\Delta^{2}+\Omega_{\textrm{p}}^{2}+\Omega_{\textrm{c}}^{2}}-|\Delta|),
\end{equation}
where $\Delta$ is the single-photon detuning of the control and probe beams from the excited state, $\Omega_{\textrm{p}}$ is the probe beam Rabi frequency, and $\Omega_{\textrm{c}}$ is the control beam Rabi frequency. When $\delta$ is tuned to the vibrational frequency $f$ of the harmonic potential by the Rabi frequencies of the control and probe beams, the cooling process takes place. For a large ensemble in an optical lattice with different trapping frequencies, the mode profiles of the cooling beams can, in principle, be tailored to match $\delta$ with $f$, while the carrier transitions are strongly suppressed for all $f$. The cooling cycle is completed when atoms decay back to the ground states in the Lamb-Dicke regime, which preserves the vibrational quantum number. In dark-state cooling, both ground states are involved in the cooling process, while in conventional Raman sideband cooling, only one ground state is used for the cooling.

\begin{figure}[h]
\subfigure{\label{fig:2a}\includegraphics[scale=0.7]{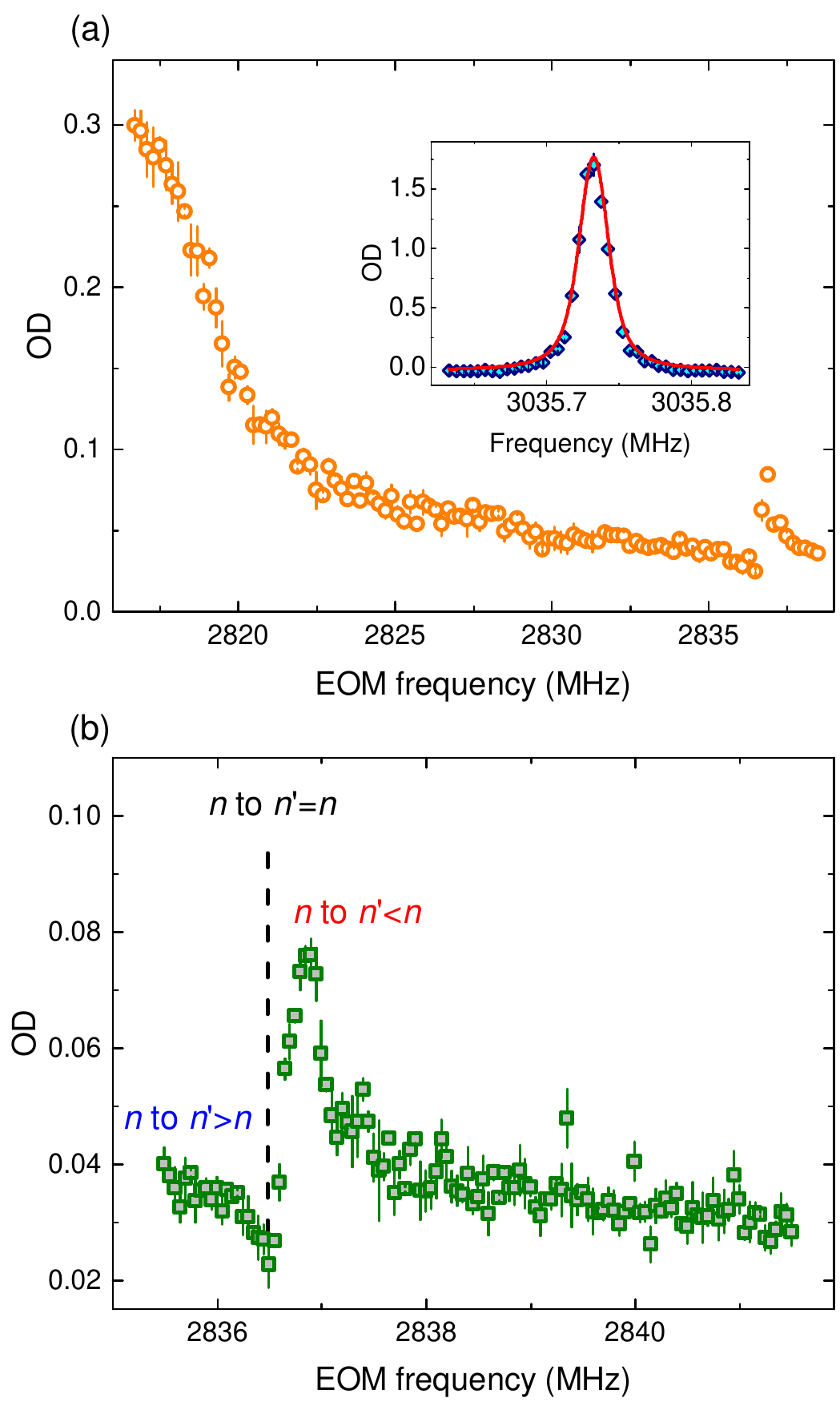}}
\caption{(a) Measurements of the probe beam absorption as a function of the probe beam frequency through tuning the EOM frequency. The absorption is characterized by the optical depth (OD). The control beam is fixed at +20 MHz detuned from the $F'$=3 state. The inset shows the Zeeman spectrum of the $F$=2 to the $F$=3 states after minimizing the ambient magnetic field. The absorption peaks from different Zeeman states collapse into a single peak. The curve is a Voigt function fitted to the experimental data with a full width at half maximum (FWHM) of 25 kHz. (b) Measurements of the probe beam absorption spectrum as a function of the probe beam EOM frequency from 2835 to 2842 MHz in (a) with finer resolution. The dashed line indicates the suppressed carrier transitions. To the right and left sides of the dashed line are red sideband and blue sideband transitions, respectively. Each data point is an average of five experimental runs, and the bars represent the standard errors.}
\label{fig:2}
\end{figure}

In our experiment, the optical lattice is formed by three co-planar free-running laser beams separated by 120$^{\circ}$, and the polarization of the lattice beams is perpendicular to the lattice plane, as shown in Fig. 1(a). The lattice potential can be written as

\begin{equation}
\begin{aligned}
U=&-\frac{4}{3}u\bigg[\frac{3}{2}+\cos(\sqrt{3}kx)+\cos\left(\frac{\sqrt{3}kx-3ky}{2}\right) \\
&+\cos\left(\frac{\sqrt{3}kx+3ky}{2}\right)\bigg],
\end{aligned}
\end{equation}
where \textit{u} is the single beam potential, \emph{k} is the wavenumber of the light, and $x$ and $y$ are the spatial coordinates. When the trap depth is large compared to the recoil energy, the potential can be approximated as a harmonic potential around $x$=0 and $y$=0 as

 \begin{equation}
U\approx-u[-6+3k^{2}(x^{2}+y^{2})],
\end{equation}
with oscillation frequency $f=\frac{1}{2\pi}\sqrt{\frac{6uk^{2}}{m}}$, where $m$ is the mass of the particle. Each lattice beam has a 1/$e^{2}$ waist of 3.5 mm with 90-mW power. The frequency of the lattice beams is 12 GHz red-detuned from the $D_{2}$ line, and the corresponding peak vibrational frequency is $f=70$ kHz \cite{Hua}. The control and probe beams are generated from the same extended cavity diode laser, whose frequency is locked to the crossover peak between the $F=3$ to $F'=2$ and $F=3$ to $F'=3$ transitions. The control beam passes through a double-pass acoustic-optical modulator (AOM) for switching and frequency-shifting of approximately +200 MHz. The probe beam passes through another double-pass AOM at approximately +400 MHz, followed by an electro-optical modulator (EOM). The positive first sideband of the EOM transmits through a passive optical cavity for the experiment. The frequency of the EOM governs the detuning of the probe beam.

The probe beam overlaps with one of the lattice beams and has a 1/$e^{2}$ waist of 2 mm, and the control beam with a 1/$e^{2}$ waist of 3 mm is aligned at 15$^{\circ}$ from the probe beam and retro-reflected to balance the radiation pressure. The polarizations of the probe and control beams are linear and perpendicular to each other. The Lamb-Dicke parameter along the $y$-axis $\eta_{y}$=$(k_{\textrm{c}}\cos 45^{\circ}+k_{\textrm{p}}\cos 60^{\circ})\sqrt{\frac{\hbar}{4\pi mf}}$ is 0.23, where $k_{\textrm{c}}$ and $k_{\textrm{p}}$ are the wavenumbers of the control and probe beams and $\cos 45^{\circ}$ and $\cos 60^{\circ}$ account for the angles between the cooling beams and the cooling axis $y$.

An ensemble of cold $^{85}$Rb atoms at 20 $\mu$K is prepared by switching on the magneto-optical trap (MOT) for 1 s succeeded by PGC on the $D_{2}$ line. The three-level $\Lambda$-type system consists of hyperfine ground states $F=2$ and $F=3$, and the excited state $F'=3$ of the $D_{1}$ line, as shown in Fig. 1(b). The sequence of the dark-state sideband cooling is shown in Fig. 1(c). The optical lattice is turned on 550 $\mu$s before the control beam. The control beam is on for 150 $\mu$s to ensure all the atoms are populated in the $F=2$ state. After a few milliseconds of the cooling sequence, the intensity of the probe beam is gradually reduced to allow atoms to accumulate in the $F=2$ state. The lattice beam intensity is then ramped off adiabatically in a few hundred microseconds.

To reduce the broadening of the hyperfine ground states from the Zeeman splitting, we apply three-axis compensating magnetic fields to minimize the ambient magnetic field. The residual magnetic field is characterized by microwave spectroscopy. We apply a microwave pulse of 0.5 ms to transfer atoms from the $F=2$ to $F=3$ states and detect the population on the $F=3$ state through absorption detection. The inset in Fig. 2(a) shows the population of atoms in the $F=3$ state as a function of microwave frequency. The full width at half maximum of the Voigt fitting function implies a residual magnetic field of 10 mG over 1 mm of the atomic cloud.

\begin{figure}[h]
\subfigure{\label{fig:2a}\includegraphics[scale=0.37]{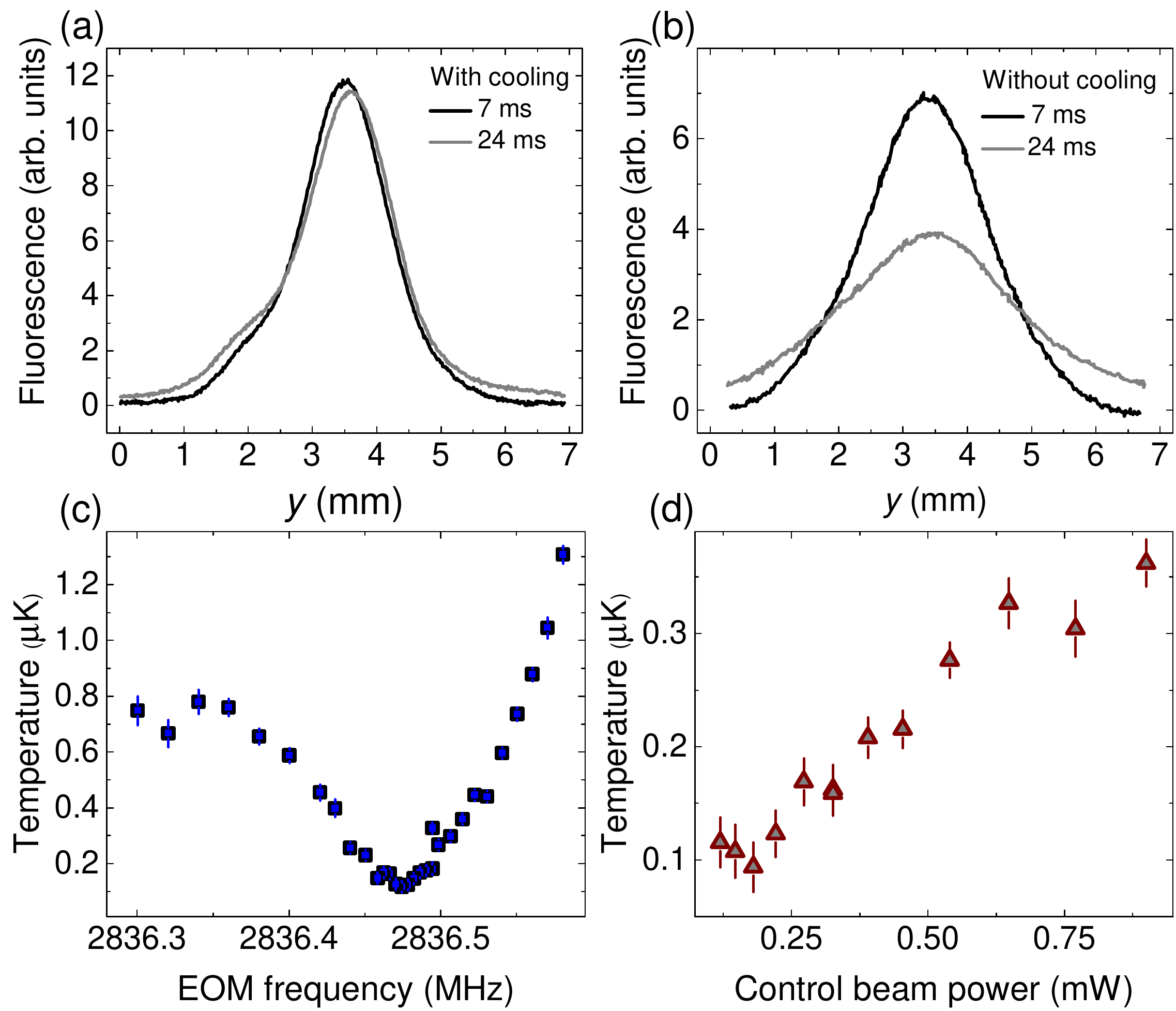}}
\caption{(a) Time-of-flight images projected on the $y$-axis at 7 and 24 ms after releasing the atoms from the optical lattice. The images are from the data point with EOM frequency of 2836.522 MHz in (c). (b)Time-of-flight images projected on the $y$-axis at the same times as in (a) without any sideband cooling. (c) Temperature measurements with varying probe beam detuning through tuning the EOM frequency. The control beam is fixed at +20 MHz detuned from the $F'$=3 state. (d) Temperature measurements with varying control beam power. The probe beam power is 0.15 mW. Each data point in (c) and (d) is an average of 20 experimental runs, and the bars represent the standard errors. The systematic error from the magnification of the imaging method is not included.}
\label{fig:2}
\end{figure}

The Fano-type absorption spectrum is measured for illustration by scanning the probe beam frequency through the EOM, as shown in Fig. 2(a). The control beam is set at +20 MHz from the resonance at 0.66 mW, and the probe beam is 0.34 mW. The relatively narrow transmission peak of the asymmetric spectrum is magnified in Fig. 2(b). The dip corresponds to the dark-state where the carrier transitions are minimized while the small peak allows red sideband transitions over a range of lattice frequencies which is determined by the width of the peak.

The temperature of the atomic ensemble is measured by the time-of-flight (TOF) method, where the two-dimensional fluorescence images of the atomic cloud are recorded 7 and 24 ms after the atoms are released from the optical lattice. We use the MOT beams to irradiate the atoms, and the fluorescence is collected by a camera along the $z$-axis. The systematic error from the uncertainty of the magnification of the imaging system on the temperature measurements is estimated to be about 20\% (10\% estimated error of the imaging system). For each two-dimensional image of the atomic cloud, we sum up the fluorescence photons on the camera pixels along the $x$-axis to obtain the distribution of the atomic cloud on the $y$-axis, as shown in Figs. 3(a) and 3(b). We observe two distributions corresponding to different temperatures. The hotter one is mainly due to the imprecision of our cooling beams' alignment with the atomic cloud. We fit the experimental data with two Gaussian functions, and the data presented throughout the rest of the article are the results of the colder atomic cloud. The fitted Gaussian widths are used to determine the temperature of the atomic cloud. The position shift of the 24-ms image from the 7-ms image is due to the angle between the line-of-sight of the camera and gravity. For comparison, we plot the distributions of the atoms after PGC without any sideband cooling in Fig. 3(b).

Figure 3(c) shows the temperature of the atoms along the $y$-axis as a function of the two-photon detuning with the single-photon detuning of the control beam set at +20 MHz. The power of the control beam is 0.33 mW, which shifts the narrow absorption peak by $\delta=140$ kHz, approximately two times the peak vibrational frequency of the optical lattice. The best cooling result is observed at two-photon resonance, which agrees with the dip of the absorption profile shown in Fig. 2(b).

While the dark state sideband cooling of atoms to the vibrational ground state has been demonstrated in various systems, the temperature $T$ of the atoms that can be achieved in combination with other cooling methods has never been explored. Here, we include adiabatic cooling after sideband cooling by switching off the optical lattice beam gradually. To extract only the dark-state sideband cooling performance, the mean vibrational number $\overline{n}$ after cooling is calculated by an equation that describes the expansion of the atoms from tight-binding bands into the free-particle band \cite{Kas}

\begin{equation}
\frac{1}{2}k_{\textrm{B}}T=E_{\textrm{R}}(\frac{Q}{k})^{2}\frac{1+4f_{\textrm{B}}+f_{\textrm{B}}^{2}}{12(1-f_{\textrm{B}})^{2}},
\end{equation}
where $k_{\textrm{B}}$ is the Boltzmann constant, $E_{\textrm{R}}$ is the recoil energy, $Q=3k/2$ is the lattice constant along the $y$-axis, and $f_{\textrm{B}}=\frac{\overline{n}}{1+\overline{n}}$ is the Boltzmann factor. For $T=100$(20) nK, the mean vibrational number $\overline{n}$ is 0.07(4), while $\overline{n}=6$ before dark-state sideband cooling. We note that $\overline{n}$ is an average over different lattice sites with different trapping frequencies. Using Eq.(5) of Ref. \cite{Mor}, the cooling-beams peak Rabi frequencies, the lattice vibrational frequency, and the cooling-beams single-photon detuning, we calculate the theoretical mean vibrational number $\overline{n}_{\textrm{th}}$=0.01. As the cooling is expected to be more efficient at the center than at the side of the atomic ensemble, our experimental result agrees reasonably well with the theory.

We vary the control beam power to shift the absorption peak and observe that the temperature decreases with smaller $\delta$, as shown in Fig. 3(d). This implies that the dark-state sideband cooling is also favorable for atoms in the potential with small vibrational frequency. When $\delta$ decreases, the absorption probability of atoms with small $f$ increases, which results in more efficient cooling. Although the absorption probability decreases for atoms with large $f$, it is still relatively large compared to the heating transition. As a result, the cooling occurs for a wide range of $f$.

\begin{figure}[h]
\subfigure{\label{fig:2a}\includegraphics[scale=0.37]{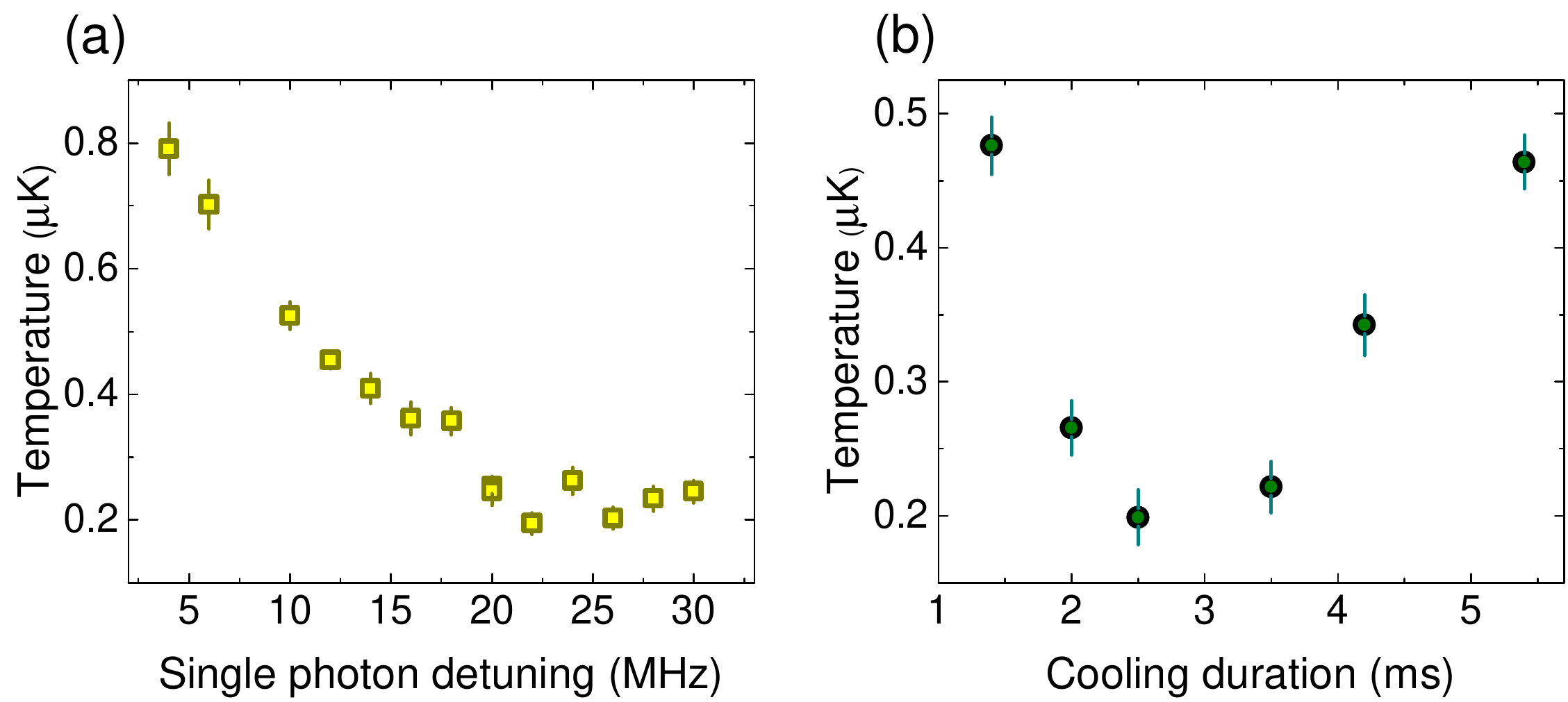}}
\caption{Temperature measurements as a function of (a) single-photon detuning and (b) cooling duration. Each data point is an average of 20 experimental runs, and the bars represent the standard errors. The systematic error from the magnification of the imaging method is not included.}
\label{fig:2}
\end{figure}

We also measure the temperature as a function of the single-photon detuning with fixed control beam power, as shown in Fig. 4(a). The result follows the trend of Eq. (1), where the increase of the single-photon detuning reduces the frequency shift. Figure 4(b) presents the temperature as a function of cooling time. The increase of the temperature after 3 ms is mainly due to the single-photon scattering from the optical lattice beams at a rate of 10$^{3}$ s$^{-1}$. As the third dimension of the temperature of the atoms is not cooled by the dark-state sideband cooling, heating due to single-photon scattering from the lattice beams in the third dimension pushes the atoms away from the center of the two-dimensional lattice, which decreases the cooling efficiency.With an optical depth of 1 using isotropic polarization of the probe beam and a size ensemble along the probe beam direction of 2 mm, the atom density is $N$=4$\times$10$^{9}$ cm$^{-3}$. The corresponding phase-space density at a temperature of 100 nK is $N\lambda_{\textrm{dB}}^{3}= 8\times10^{-4}$, where $\lambda_{\textrm{dB}}$ is the thermal de Broglie wavelength.

In summary, we demonstrated cooling of atoms to the vibrational ground state in an optical lattice and sub-recoil temperature with the aid of adiabatic cooling within a few milliseconds. Our results are in good qualitative agreement with the theoretical predictions for the explored parameter ranges. The lowest temperature achieved, in units of the recoil temperature, is a factor of 4 lower than the previous result \cite{Ker}. In addition, the experimental implementation is relatively simple compared to other sideband cooling schemes. This method can be applied to atoms trapped in two- or three-dimensional optical lattices whose trapping frequencies vary across different lattice sites due to the Gaussian distribution of the laser beams. In particular, the dark-state sideband cooling would be advantageous for atoms trapped in lattices formed by optical waveguides \cite{Xin,Men}, where the intensity distribution of the lattice beams follows the waveguide modes. The fast cooling time and cooling from tens of micro-kelvin to sub-recoil temperature could be used as a pre-cooling stage to produce a large ensemble of ultra-cold atoms or molecules \cite{Ande} to study ultra-cold physics.

This work is supported by the Singapore National Research Foundation under Grant No. NRFF2013-12 and QEP-P4, and the Singapore Ministry of Education under Grant No. MOE2017-T2-2-066.


\nocite{*}

\bibliography{apssamp}

\begin{thebibliography}{00}
\bibitem{Mor}G. Morigi, J. Eschner, and C. H. Keitel, Ground State Laser Cooling Using Electromagnetically Induced Transparency, Phys. Rev. Lett. {\bf85}, 4458 (2000).
\bibitem{Mor2}G. Morigi, Cooling Atomic Motion with Quantum Interference, Phys. Rev. A {\bf67}, 033402 (2003).
\bibitem{Met}H. Metcalf and P. van der Straten, Laser Cooling and Trapping (Springer, 1999).
\bibitem{And}M. H. Anderson, J. R. Ensher, M. R. Matthews, C. E. Wieman, E. A. Cornell, Observation of Bose-Einstein Condensation in a Dilute Atomic Vapor, Science {\bf269}, 198 (1995).
\bibitem{Kov}T. Kovachy, J. M. Hogan, A. Sugarbaker, S. M. Dickerson, C. A. Donnelly, C. Overstreet, and M. A. Kasevich, Matter Wave Lensing to Picokelvin Temperatures, Phys. Rev. Lett. {\bf114}, 143004 (2015).
\bibitem{Kase} M. Kasevich and S. Chu, Laser Cooling Below a Photon Recoil with Three-Level Atoms. Phys. Rev. Lett. {\bf69}, 1741 (1992).
\bibitem{Per}H. Perrin, A. Kuhn, I. Bouchoule, and C. Salomon, Sideband Cooling of Neutral Atoms in a Far-Detuned Optical Lattice, Europhys. Lett. {\bf42}, 395 (1998).
\bibitem{Ham} S. E. Hamann, D. L. Haycock, G. Klose, P. H. Pax, I. H. Deutsch, and P. S. Jessen, Resolved-Sideband Raman Cooling to the Ground State of an Optical Lattice, Phys. Rev. Lett. {\bf80}, 4149 (1998).
\bibitem{Vul}V. Vuleti\'{c}, C. Chin, A. J. Kerman, and S. Chu, Degenerate Raman Sideband Cooling of Trapped Cesium Atoms at Very High Atomic Densities, Phys. Rev. Lett. {\bf81}, 5768 (1998).
\bibitem{Ker}A. J. Kerman, V. Vuleti\'{c}, C. Chin, and S. Chu, Beyond Optical Molasses: 3D Raman Sideband Cooling of Atomic Cesium
to High Phase-Space Density. Phys. Rev. Lett. {\bf84}, 439 (2000).
\bibitem{Asp}A. Aspect, E. Arimondo, R. Kaiser, N. Vansteenkiste, and C. Cohen-Tannoudji, Laser Cooling below the One-Photon Recoil Energy by Velocity-Selective Coherent Population Trapping, Phys. Rev. Lett. {\bf61}, 826 (1988).
\bibitem{Roo}C. F. Roos, D. Leibfried, A. Mundt, F. Schmidt-Kaler, J. Eschner, and R. Blatt, Experimental Demonstration of Ground State Laser Cooling with Electromagnetically Induced Transparency, Phys. Rev. Lett. {\bf85}, 5547 (2000).
\bibitem{Lin}Y. Lin, J. P. Gaebler, T. R. Tan, R. Bowler, J. D. Jost, D. Leibfried, and D. J. Wineland, Sympathetic Electromagnetically-Induced-Transparency Laser Cooling of Motional Modes in an Ion Chain. Phys. Rev. Lett. {\bf110}, 153002 (2013).
\bibitem{Lec}R. Lechner, C. Maier, C. Hempel, P. Jurcevic, B. P. Lanyon, T. Monz, M. Brownnutt, R. Blatt, and C. F. Roos, Electromagnetically-Induced-Transparency Ground-State Cooling of Long Ion Strings, Phys. Rev. A {\bf93}, 053401 (2016).
\bibitem{Sch}N. Scharnhorst, J. Cerrillo, J. Kramer, I. D. Leroux, J. B. W\"{u}bbena, A. Retzker, and P. O. Schmidt, Experimental and Theoretical Investigation of a Multimode Cooling Scheme Using Multiple Electromagnetically-Induced-Transparency Resonances, Phys. Rev. A {\bf98}, 023424 (2018).
\bibitem{Jor}E. Jordan, K. A. Gilmore, A. Shankar, A. Safavi-Naini, J. G. Bohnet, M. J. Holland, and J. J. Bollinger, Near Ground-State Cooling of Two-Dimensional Trapped-Ion Crystals with More than 100 Ions, Phys. Rev. Lett. {\bf122}, 053603 (2019).
\bibitem{Qia}M. Qiao, Y. Wang, Z. Cai, B. Du, P. Wang, C. Luan, W. Chen, H.-R. Noh, and K. Kim, Double-EIT Ground State Cooling of Stationary Two-Dimensional Ion Lattices, arXiv:2003.10276 (2020).
\bibitem{Fen}L. Feng, W. L. Tan, A. De, A. Menon, A. Chu, G. Pagano, and C. Monroe, Efficient Ground-State Cooling of Large Trapped-Ion Chains with an EIT Tripod Scheme, Phys. Rev. Lett. {\bf125}, 053001 (2020).
\bibitem{Kam}T. Kampschulte, W. Alt, S. Manz, M. Martinez-Dorantes, R. Reimann, S. Yoon, D. Meschede, M. Bienert, and G. Morigi, Electromagnetically-Induced-Transparency Control of Single-Atom Motion in an Optical Cavity. Phys. Rev. A {\bf89}, 033404 (2014).
\bibitem{Hal}E. Haller, J. Hudson, A. Kelly, D. A. Cotta, B. Peaudecerf, G. D. Bruce, and S. Kuhr, Single-Atom Imaging of Fermions in a Quantum-Gas Microscope, Nature Physics {\bf11}, 738 (2015).
\bibitem{Edg}G. J. A. Edge, R. Anderson, D. Jervis, D. C. McKay, R. Day, S. Trotzky, and J. H. Thywissen, Imaging and Addressing of Individual Fermionic Atoms in an Optical Lattice, Phys. Rev. A {\bf92}, 063406 (2015).
\bibitem{Lou}B. Lounis and C. Cohen-Tannoudji, Coherent Population Trapping and Fano Profiles, J. Phys. II (France) {\bf2}, 579 (1992).
\bibitem{Hua}C. Huang, P.-C. Kuan, and S.-Y. Lan, Laser Cooling of $^{85}$Rb Atoms to the Recoil-Temperature Limit, Phys. Rev. A {\bf97}, 023403 (2018).
\bibitem{Kas}A. Kastberg, W. D. Phillips, S. L. Rolston, R. J. C. Spreeuw, and P. S. Jessen, Adiabatic Cooling of Cesium to 700 nK in an Optical Lattice, Phys. Rev. Lett. {\bf74}, 1542 (1995).
\bibitem{Xin}M. Xin, W. S. Leong, Z. Chen, and S.-Y. Lan, Transporting Long-Lived Quantum Spin Coherence in a Photonic Crystal Fiber. Phys. Rev. Lett. {\bf122}, 163901 (2019).
\bibitem{Men}Y. Meng, A. Dareau, P. Schneeweiss, and A. Rauschenbeutel, Near-Ground-State Cooling of Atoms Optically Trapped 300 nm Away From a Hot Surface. Phys. Rev. X {\bf8}, 031054 (2018).
\bibitem{Ande} L. Anderegg, B. L. Augenbraun, Y. Bao, S. Burchesky, L. W. Cheuk, W. Ketterle, and J. M. Doyle, Laser Cooling of Optically Trapped Molecules. Nat. Phys. {\bf14}, 890 (2018).


\end{thebibliography}

\end{document}